\documentclass[a4paper]{article}

\usepackage{amssymb}
\usepackage{graphicx}
\usepackage{geometry}
\usepackage[cm]{fullpage}
\usepackage{xcolor}
\usepackage[hidelinks]{hyperref}

\newcommand{\figureref}[1]{{\color{black} Fig.~#1}}
\newcommand{\todo}[1]{{\color{black}#1}}

\begin{document}

\title{Coherent spin-exchange via a quantum mediator}
\author{T.~A. Baart$^{1}$\footnote{Contributed equally}, T. Fujita$^{1*}$, C. Reichl$^{2}$,\\ W. Wegscheider$^{2}$, L.~M.~K. Vandersypen$^{1}$\footnote{email: l.m.k.vandersypen@tudelft.nl}}
\maketitle

\begin{enumerate}
	\item QuTech and Kavli Institute of Nanoscience, TU Delft, 2600 GA Delft, The Netherlands
	\item Solid State Physics Laboratory, ETH Z\"{u}rich, 8093 Z\"{u}rich, Switzerland
\end{enumerate}

\textbf{Coherent interactions at a distance provide a powerful tool for quantum simulation and computation. The most common approach to realize an effective long-distance coupling `on-chip' is to use a quantum mediator, as has been demonstrated for superconducting qubits~\cite{Majer2007,Sillanpaa2007} and trapped ions~\cite{Schmidt-Kaler2003}. For quantum dot arrays, which combine a high degree of tunability~\cite{Hanson2007} with extremely long coherence times~\cite{Veldhorst2015}, the experimental demonstration of coherent spin-spin coupling via an intermediary system remains an important outstanding goal~\cite{Taylor2006,Burkard2006,Hu2012,Schuetz2015,Trifunovic2012,Trifunovic2013,Friesen2007,Leijnse2013,Hassler2015,Trauzettel2007,Frey2012,Mehl2014,Lehmann2007,Stano2015,Srinivasa2015,Viennot2015,Petersson2012}. Here, we use a linear triple-quantum-dot array to demonstrate a first working example of a coherent interaction between two distant spins via a quantum mediator. The two outer dots are occupied with a single electron spin each and the spins experience a superexchange interaction through the empty middle dot which acts as mediator. Using single-shot spin read-out~\cite{Elzerman2004a} we measure the coherent time evolution of the spin states on the outer dots and observe a characteristic dependence of the exchange frequency as a function of the detuning between the middle and outer dots. This approach may provide a new route for scaling up spin qubit circuits using quantum dots and aid in the simulation of materials and molecules with non-nearest neighbour couplings such as MnO~\cite{Kramers1934,Anderson1950}, high-temperature superconductors~\cite{Kim1998,Kohno2014} and DNA~\cite{Giese2001}. The same superexchange concept can also be applied in cold atom experiments~\cite{Meinert2014}.}\\

Nanofabricated quantum dot circuits provide an excellent platform for performing both quantum computation and simulation using single spins~\cite{Hanson2007,Loss1998a,Taylor2005,Barthelemy2013}. Many approaches to implementing coherent spin coupling between distant quantum dots have been proposed using a variety of coupling mechanisms. These include superconducting resonators~\cite{Taylor2006,Burkard2006,Hu2012}, surface-acoustic wave resonators~\cite{Schuetz2015}, floating metallic~\cite{Trifunovic2012} or ferromagnetic couplers~\cite{Trifunovic2013}, collective modes of spin chains~\cite{Friesen2007}, superconductors~\cite{Leijnse2013,Hassler2015}, Klein tunneling through the valence or conduction band~\cite{Trauzettel2007} and superexchange or sequential operations via intermediate quantum dots~\cite{Mehl2014,Lehmann2007,Stano2015,Srinivasa2015}. A common theme among many of these proposals is to create a coupling between distant spins by virtual occupation of a mediator quantum system. So far, none of these schemes have been performed experimentally. More broadly, there are no experimental realizations so far of direct quantum gates between any type of solid-state spins at a distance.\\

In this Letter we focus on the superexchange interaction to induce spin-spin coupling at a distance. Superexchange is the (usually) antiferromagnetic coupling between two next-to-nearest neighbour spins through virtual occupation of a non-magnetic intermediate state~\cite{Kramers1934,Anderson1950}. To date, only indirect observations of superexchange involving spins in quantum dots have been reported~\cite{Sanchez2014b}. Given that superexchange involves a fourth order process in the hopping amplitude, it is challenging to use it for achieving coherent coupling. This is also the case for several related schemes relying on quantum mediators.\\

We use a linear triple-quantum-dot array with one electron on each of the outer dots, and induce a superexchange interaction through the empty middle dot, which acts as a quantum mediator. This induces spin exchange of the two distant electron spins. Using repeated single-shot spin measurements we record the coherent time evolution of the spin states on the outer dots. We control the superexchange amplitude via the detuning of the middle dot electrochemical potential relative to those of the outer dots, and study the cross-over between superexchange and conventional nearest-neighbour spin exchange.\\

The dot array is formed electrostatically in a two-dimensional electron gas (2DEG) 85 nm below the surface of a GaAs/AlGaAs heterostructure, see \figureref{1a}. Gate electrodes fabricated on the surface (see Methods) are biased with appropriate voltages to selectively deplete regions of the 2DEG and define the linear array of three quantum dots. The left and right dot are each occupied with one electron, and each of the two electrons constitutes a single spin-$\frac{1}{2}$ particle. The interdot tunnel couplings are set to $\approx 8.5$~GHz (left-middle) and $\approx 11.8$~GHz (middle-right). The sensing dot (SD) next to the quantum dot array is used for non-invasive charge sensing using radiofrequency (RF) reflectometry to monitor the number of electrons in each dot~\cite{Barthel2010}. An in-plane magnetic field $B_{ext}$ = 3.2 T is applied to split the spin-up ($\uparrow$) and spin-down ($\downarrow$) states of each electron by the Zeeman energy ($E_{Z} \approx 80$ $\mathrm{\mu}$eV), defining a qubit. The electron temperature of the right reservoir is $\approx 75$~mK.\\

In this system, superexchange can be seen as the result of the effective tunnel coupling $t_{SE}$ between the outer dots. The amplitude of superexchange, $J_{SE}$, is approximated by $-\frac{t_{SE}^2}{\epsilon}$, with $\epsilon$ the detuning between the electrochemical potentials of the outer dots~\cite{Loss1998a}, and $\epsilon=0$ when (1,0,1) and (2,0,0) are degenerate. Here $t_{SE}$ can be described as $t_{SE} = (t_{m,l}t_{m,r})/\delta$, with $t_{m,l}$ ($t_{m,r}$) the tunnel coupling between the middle and the left (right) site and $\delta$ the detuning between the electrochemical potential of (1,1,0) and the average of the electrochemical potentials of (1,0,1) and (2,0,0)~\cite{Braakman2013}. The superexchange amplitude can thus be approximated as (see Supplementary Information \todo{VI} for the range of validity)
\begin{equation}
J_{SE} = - \frac{t^2_{m,l}t^2_{m,r}}{\delta^2 \epsilon},
\label{eq:J_{SE}}
\end{equation}
which illustrates the characteristic fourth-order hopping process underlying superexchange.\\

To provide direct evidence of coherent superexchange, we will probe the resulting time evolution of the two spins via repeated single-shot measurements using spin-to-charge conversion~\cite{Elzerman2004a}. It is beneficial to perform this conversion as close as possible to the charge sensor, SD, to achieve high read-out fidelities. In previous work, we therefore shuttled electrons consecutively from left to middle to right with no detectable sign of spin flips upon shutting~\cite{Baart2015}. Here, we explore a different approach, transferring the spin from left to right with only virtual occupation of the middle dot, using the same long-range tunnel coupling that underlies coherent superexchange~\cite{Sanchez2014b}. We test the two-spin read-out and long-range spin transfer as described by the schematic diagrams of \figureref{1b} and implemented by the pulse sequence depicted by the blue and red arrows in \figureref{1c}. Starting from an empty array, we load a random electron from the reservoir into the right dot by pulsing into the charge state (0,0,1). Next we pulse into (1,0,0), whereby the electron is transferred from the rightmost dot to the leftmost dot via a second-order tunnel process across the middle dot. For this transfer we temporarily pulse $\delta$ closer to 0 to increase the long-range shuttling rate (see Supplementary Information \todo{II}). Finally, we once more load a random electron in the right dot by pulsing to (1,0,1). We vary the waiting time in (1,0,1) during which spins relax to the spin ground state $\left| \uparrow0\uparrow \right>$. Then we reverse the pulse sequence and add two calibrated read-out stages denoted by the green circles where spin-to-charge conversion takes place. \figureref{1d} shows the measured decays to the ground state spin-up for each of the two spins. We report read-out fidelities of on average 95.9\% and 98.0\% for spin-down and spin-up respectively, assuming no spin flips during the spin transfer~\cite{Baart2015} (see Supplementary Information \todo{IV}). \\

A key signature of superexchange driven spin oscillations is their dependence on the detuning of the intermediate level ($\delta$), see Eq.~(\ref{eq:J_{SE}}). We have therefore created linear combinations of the gates $P_{1}$, $P_{2}$ and $P_{3}$ in such a way that we can independently vary $\delta$ and $\epsilon$ as can be seen in \figureref{2b}. Superexchange occurs in the (1,0,1) charge configuration, and the superexchange amplitude, $J_{SE}$, increases for less negative $\epsilon$, which translates to an operating point closer to the (2,0,0)-configuration, see \figureref{2a}. Similarly, $J_{SE}$ increases with less negative $\delta$, up to the point where we cross the (1,0,1)-(1,1,0) transition indicated by the black dashed line in \figureref{2b} and spin exchange between nearest-neighbour dots will dominate (see \figureref{2c}). To capture the expected time evolution, we must take into account a difference in Zeeman energies between the two dots, $\Delta E_{z} = E_{z,3} - E_{z,1}$, arising from slight differences in the $g$-factor for each dot~\cite{Baart2015}. Spin exchange defines one rotation axis, the Zeeman energy difference an orthogonal axis, as shown in the Bloch sphere in \figureref{2d}. In the experiment, $\Delta E_{z}$ is fixed, and $J_{SE}$ can be controlled by gate voltage pulses, as we discussed. By adjusting $J_{SE}$, we can thus define the net rotation axis and rate~\cite{Dial2013}. A similar Bloch sphere can be made for the nearest-neighbour regime. \\

The protocol for probing the time evolution is as follows. Starting with an empty array, we create a mixture of $\left| \uparrow0\downarrow \right>$ and $\left| \uparrow0\uparrow \right>$ and move to the position of the red star in \figureref{2b}, where $J_{SE}$ is small compared to $\Delta E_Z$. This is achieved by sequentially loading the two spins as in \figureref{1c}, in this case loading a $\uparrow$ in the left dot and a random spin in the right dot. This procedure allows us to conveniently create an anti-parallel spin state without using more involved techniques such as electron spin resonance. Next, following the black dashed arrows in \figureref{1c}, we pulse towards the (2,0,0) regime and wait for several ns. The exact location in detuning space is marked in \figureref{2b} by a red diamond. At this point $J_{SE}$ is sizable, $\left| \uparrow0\downarrow \right>$ is not an eigenstate of the Hamiltonian and is thus expected to evolve in time, periodically developing a $\left| \downarrow0\uparrow \right>$ component ($\left| \uparrow0\uparrow\right>$  will only acquire an overall phase). The larger $J_{SE}/\Delta E_Z$, the larger the $\left| \downarrow0\uparrow \right>$ component. We pulse back to the position of the red star in (1,0,1) and follow the same spin read-out procedure as was done for the $T_{1}$-measurement in \figureref{1d}. \figureref{2e} shows the $\left| \uparrow0\downarrow \right>$ and $\left| \downarrow0\uparrow \right>$ probability as a function of the length of the detuning pulse. We see a sinusoidal dependence, with the $\left| \uparrow0\downarrow \right>$ and $\left| \downarrow0\uparrow \right>$ populations evolving in anti-phase, as expected.\\ 

Returning to the key signature of superexchange, we fix the value of $\epsilon$ and vary $\delta$ along the vertical dashed line shown in \figureref{2b}. For each choice of $\delta$, we record the four two-spin probabilities as a function of the length of the detuning pulse (\figureref{3a}). Starting from large negative $\delta$, we first observe no oscillations at all: the superexchange mechanism is suppressed and the  $\left| \uparrow0\downarrow \right>$-state remains fixed along the $x$-axis of the Bloch sphere. As we bring the electrochemical potential of the intermediate level closer to that of the outer dots, $J_{SE}$ increases in magnitude and slow oscillations $\sim 150$~MHz start appearing that are still dominated by $\Delta E_{z} \approx 130$~MHz between the outer dots, hence the low contrast of the oscillations. The oscillations become faster up to $\sim900$~MHz as $\delta$ is increased at which point $J_{SE}$ is stronger than $\Delta E_{z}$ and the contrast increases. When $\delta$ is further increased, the (1,1,0)-state becomes energetically favourable and the nearest-neighbour exchange between the left and middle dot dominates. Here $\epsilon = - 170~\mu$eV and this transition occurs around $\delta = 120~\mu$eV, which is where the black-dashed line in \figureref{2b} is crossed. Increasing $\delta$ even more enlarges the detuning between the left and middle dot and thereby slows down the nearest-neighbour oscillations, as seen in the data.\\ 

For a quantitative comparison with the theoretical predictions, we show in \figureref{3b} the expected time evolution of the system modeled using the measured nearest-neighbour tunnel couplings, detunings $\delta$ and $\epsilon$, and the difference in Zeeman energy probed through electric-dipole spin resonance measurements~\cite{Shafiei2013}. We include the effect of dephasing by charge noise~\cite{Dial2013} via a single parameter to match the decay of the oscillations and account for the known read-out fidelities (see Supplementary Information \todo{V}). \figureref{2e} shows that it takes more than 1~ns for the superexchange to be turned on. This is caused by the finite risetime of the pulses produced by the arbitrary waveform generator and finite bandwidth of the coax lines. The simulation includes this gradual turn on and off of $J_{SE}$. Comparing \figureref{3a} and \figureref{3b} we report good agreement between theory and experiment, which supports our interpretation of the data in terms of superexchange, including the transition to nearest-neighbour exchange.\\

In summary, we have demonstrated a first working example of a direct quantum gate between solid-state spins at a distance via virtual occupation of a quantum mediator. This result underlines the utility of arrays of quantum dots for the investigation and application of fundamental physical processes driven by small-amplitude terms and higher-order tunneling. It is possible to extend the distance between the coupled spins using elongated intermediate quantum dots or via different (quantum) mediators altogether. Another interesting direction is to create non-nearest neighbour spin-spin interactions with the centre dot occupied~\cite{Stano2015,Srinivasa2015}, which opens up further new possibilities for quantum computation and modeling of complex materials.\\

 
\newpage

\textbf{Additional information} Supplementary Information is linked to the online version of the paper.\\

\textbf{Acknowledgements} The authors acknowledge useful discussions with the members of the Delft spin qubit team, sample fabrication by F.R. Braakman, and experimental assistance from M. Ammerlaan, J. Haanstra, R. Roeleveld, R. Schouten, M. Tiggelman and R. Vermeulen. This work is supported by the Netherlands Organization of Scientific Research (NWO) Graduate Program, the Intelligence Advanced Research Projects Activity (IARPA) Multi-Qubit Coherent Operations (MQCO) Program, the Japan Society for the Promotion of Science (JSPS) Postdoctoral Fellowship for Research Abroad and the Swiss National Science Foundation.\\

\textbf{Author contributions} T.A.B and T.F. performed the experiment and analyzed the data, C.R. and W.W. grew the heterostructure, T.A.B., T.F. and L.M.K.V. contributed to the interpretation of the data, and T.A.B. and L.M.K.V. wrote the manuscript, with comments from T.F. \\

\textbf{Author information} The authors declare no competing financial interests. Correspondence and requests for materials should be addressed to L.M.K.V.\\
 
\newpage

\noindent \textbf{Figure 1 Linear array of three quantum dots and long-range spin transfer}\\
\textbf{a} Scanning electron microscopy image of a sample nominally identical to the one used for the measurements. Dotted circles indicate quantum dots and squares indicate Fermi reservoirs in the 2DEG, which are connected to ohmic contacts. The RF reflectance of the SD is monitored in order to determine the occupancies of the three dots labeled numbers 1 to 3 from left to right respectively. \textbf{b} Read from left to right and top to bottom. The array is initialized by loading two electrons from the right reservoir. The spin that is loaded first is transferred to the left dot via a second-order tunnel process across the middle dot. We load $\uparrow$-spins by tuning the loading position such that only the $\uparrow$-spin level is accessible (as in the top left diagram). Random spins are loaded by making both spin levels energetically available (top right). Spin read-out occurs using energy-selective tunneling combined with charge detection via the SD. \textbf{c} Charge stability diagram of the triple dot for $M=$~-412 mV. Along the $L$ and $R$ axis, we linearly vary the voltages applied to gates $P_{1}$, $P_{2}$ and $P_{3}$ in such a way that we affect mostly the left and right dots, compensating for cross-capacitances. Similarly, $M$ controls mostly the middle dot (see Supplementary Information \todo{III}). Labels $(n,m,p)$ indicate the number of electrons in the left, middle and right dot respectively. The middle dot cannot be loaded directly from a reservoir~\cite{Yang2014} and the left dot is only weakly tunnel coupled to the left reservoir, leading to faintly visible charge transitions (black dotted lines indicate their positions). The pulse sequence for loading and read-out is indicated in the charge stability diagrams via blue and red arrows, see also panel b. The two black dashed arrows denote additional stages to probe superexchange (see \figureref{2}). \textbf{d} Measured single-spin populations averaged over 8000 cycles per datapoint as a function of waiting time in (1,0,1) for dot 1 (top) and dot 3 (bottom).

\newpage

\noindent \textbf{Figure 2 Superexchange-driven spin oscillations}\\
\textbf{a} Energy diagram as a function of $\epsilon$ for $\delta<0$. The long-range tunnel coupling induces an anti-crossing between the (1,0,1) and (2,0,0) singlet states. The energy difference between $T_{0}$ and the hybridized $S$ is denoted $J_{SE}$. The $T_{-}$ and $T_{+}$ states are split off by $B_{ext}$. \textbf{b} Charge stability diagram in detuning space, allowing individual control of the detuning of the middle dot ($\delta$) and between the outer dots ($\epsilon$), see panel c. \textbf{c} Cartoon depicting the transition from superexchange to nearest-neighbour exchange as $\delta$ is made more positive. \textbf{d} Bloch sphere representation of $S-T_0$ subspace in the superexchange regime with control axes $J_{SE}$ and $\Delta E_{Z}$. \textbf{e} Observation of superexchange-driven spin oscillations. Starting with a mixture of $\left| \uparrow 0 \downarrow \right> $ and $\left| \uparrow 0 \uparrow \right>$ at the position of the red star in b, we pulse $\epsilon$ for a varying amount of time to the position indicated by the red diamond. Afterwards the four two-spin probabilities are measured by averaging over 999 single-shot cycles per datapoint, two of which are shown.  

\newpage

\noindent \textbf{Figure 3 Transition from superexchange to nearest-neighbour exchange}\\
\textbf{a} Starting with a mixture of $\left| \uparrow 0 \downarrow \right> $ and $\left| \uparrow 0 \uparrow \right>$ at the position of the red star in \figureref{2b}, we pulse $\epsilon$ and $\delta$ for a varying amount of time to the position indicated by the vertical dashed line in \figureref{2b}. Afterwards the four two-spin probabilities are measured by averaging over 999 single-shot cycles per datapoint. We clearly note the transition of oscillations dominated by $\Delta E_{z}$ ($\delta < -50~\mu$eV) to increasingly faster superexchange dominated spin evolution and finally ($\delta>200~\mu$eV) nearest-neighbour exchange dominated evolution, which slows down as $\delta$ is further increased. Acquiring this set of data took $\sim$20 hours. \textbf{b} Simulation of the data shown in a. The independently determined input parameters are: $t_{m,l}=8.5$~GHz, $t_{m,r}=$11.8~GHz,  $E_{z,1}=$19.380~GHz, $E_{z,2}=$19.528~GHz, $E_{z,3}=$19.510~GHz and the risetime of the detuning pulse is 0.8~ns (see Supplementary Information \todo{V}).

\newpage

\begin{figure*}[h!]
	\centering
	\includegraphics[width=1.0\textwidth]{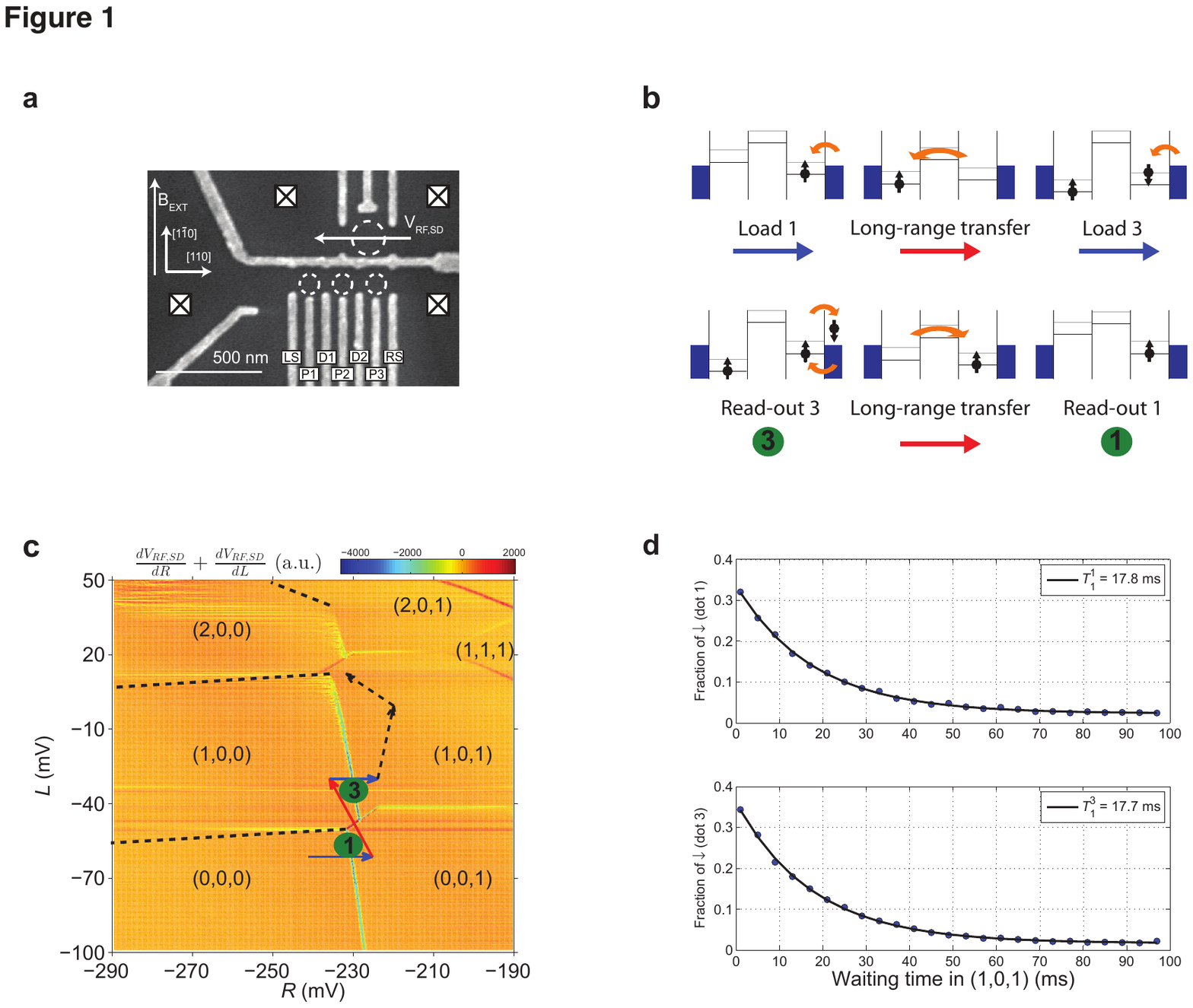}
\end{figure*}

\newpage

\begin{figure*}[h!]
	\centering
	\includegraphics[width=1.0\textwidth]{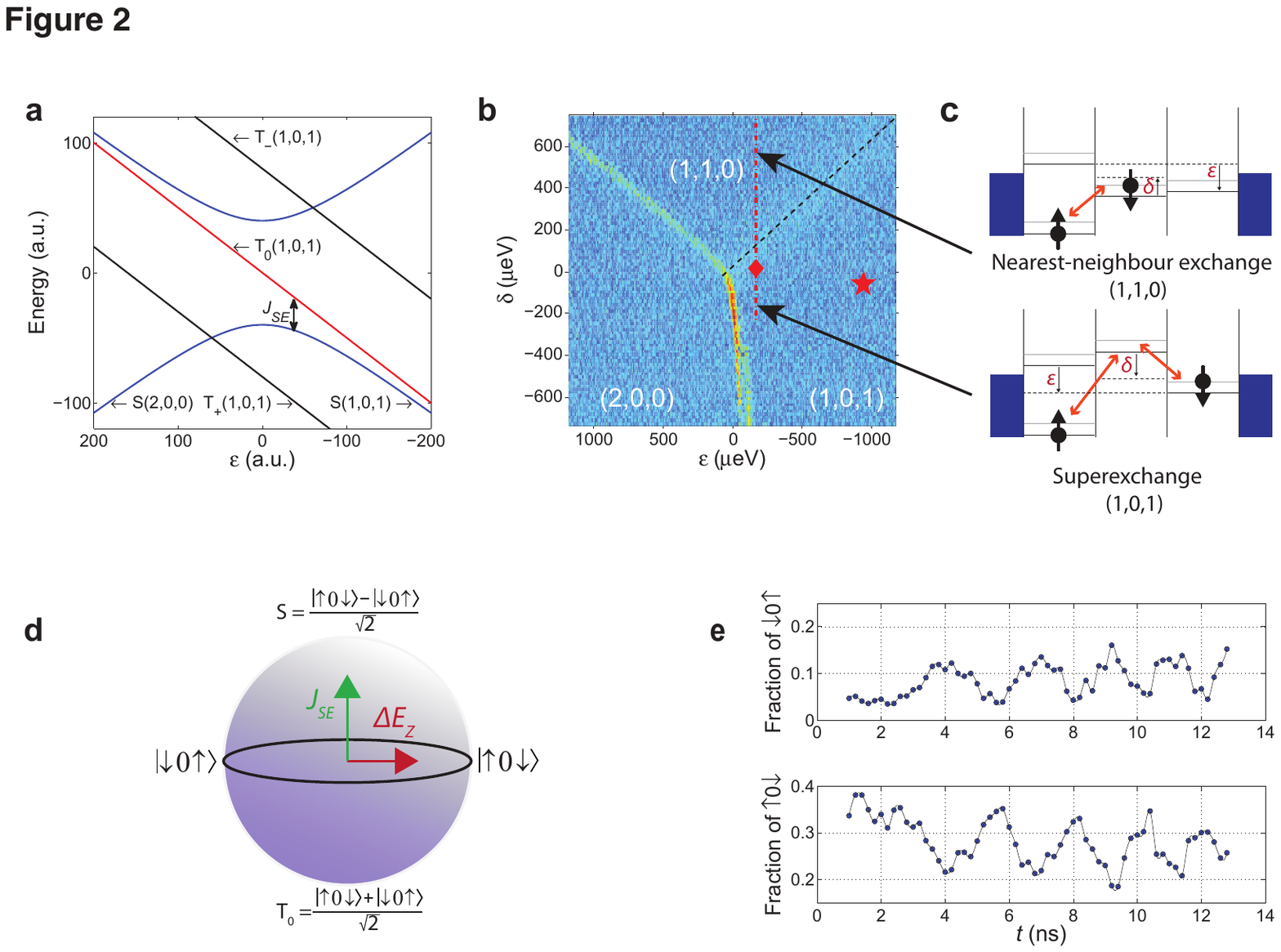}
\end{figure*}

\newpage

\begin{figure*}[h!]
	\centering
	\includegraphics[width=1.0\textwidth]{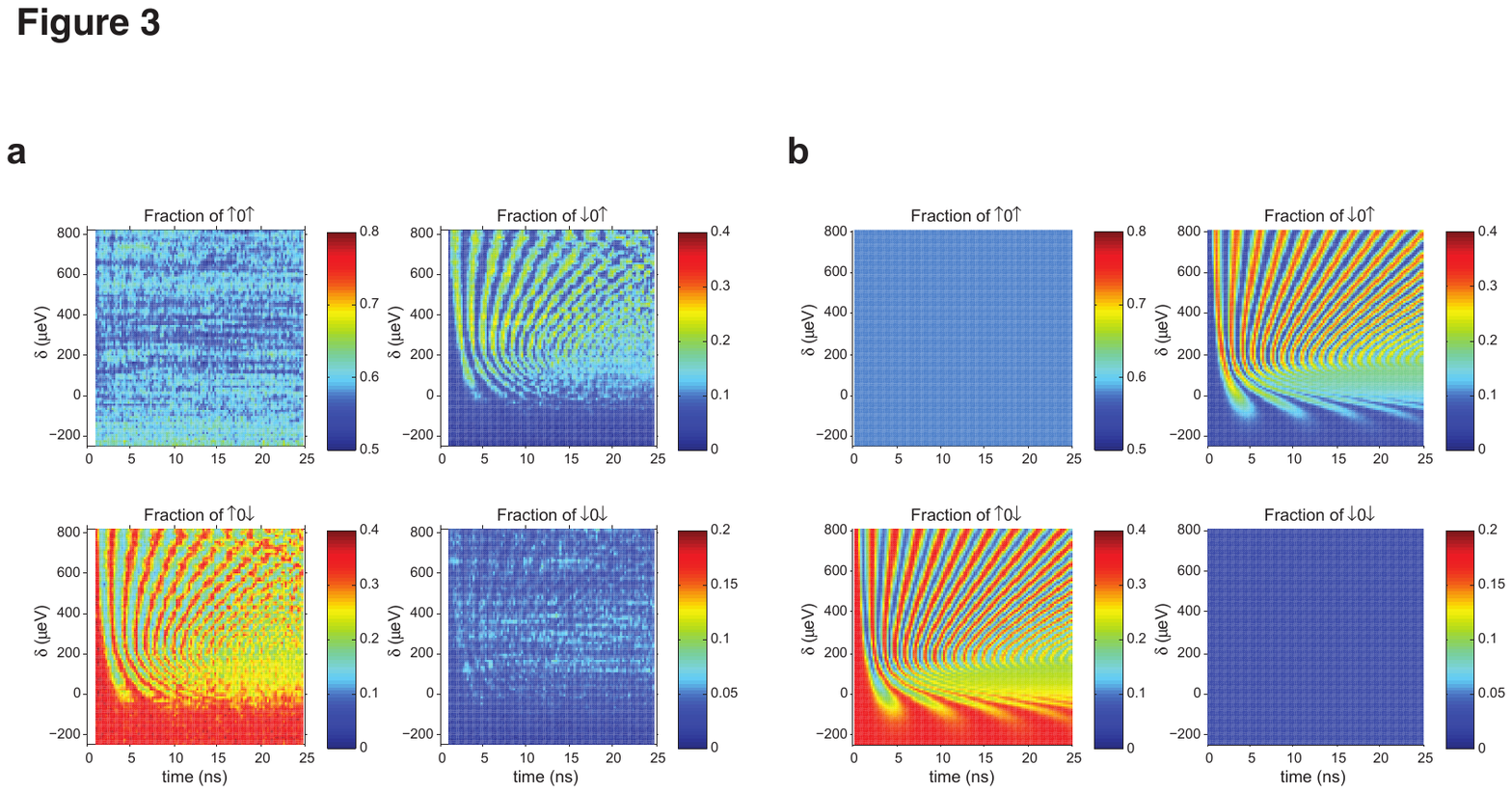}
\end{figure*}

\clearpage
\newpage

\renewcommand{\figurename}{Fig. S}
\renewcommand{\tablename}{Table S}
\renewcommand{\thesection}{\Roman{section}} 
\renewcommand{\thesubsection}{\Roman{subsection}}

	\begin{centering}
		{\Large Supplementary Information for} \\ \vspace{0.2cm}
		{\Large \textbf{Coherent spin-exchange via a quantum mediator}}\\
		\vspace{0.4cm}
		
		{\normalsize T.~A. Baart$^{1}$, T. Fujita$^{1}$,}\\  
		{\normalsize C. Reichl$^{2}$, W. Wegscheider$^{2}$, L.~M.~K. Vandersypen$^{1}$}\\
		\vspace{0.4cm}
		\normalsize{$^{1}$QuTech and Kavli Institute of Nanoscience, TU Delft, 2600 GA Delft, The Netherlands}\\
		\normalsize{$^{2}$Solid State Physics Laboratory, ETH Z\"{u}rich, 8093 Z\"{u}rich, Switzerland}\\
	\end{centering}
	
	\tableofcontents
	
	\newpage

Parts of the supplementary material were already published in \cite{Baart2015_sup} and are repeated here to make the text more self-contained.
	
\section{Methods and materials}
The experiment was performed on a $\mathrm{GaAs/Al_{0.25}Ga_{0.75}As}$ heterostructure grown by molecular-beam epitaxy, with a 85-nm-deep 2DEG with an electron density of $\mathrm{2.0 \cdot 10^{11}\ cm^{-2}}$ and mobility of $\mathrm{5.6 \cdot 10^{6}\ cm^{2} V^{-1} s^{-1}}$ at 4 K. The metallic (Ti-Au) surface gates were fabricated using electron-beam lithography. The device was cooled inside an Oxford Kelvinox 400HA dilution refrigerator to a base temperature of 45 mK. To reduce charge noise, the sample was cooled while applying a positive voltage on all gates (ranging between 250 and 350 mV)~\cite{Long2006_sup}. The main function of gates $LS$ and $RS$ is to set the tunnel coupling with the left and right reservoir, respectively. $D_{1}$ and $D_{2}$ control the interdot tunnel coupling and $P_{1}$, $P_{2}$ and $P_{3}$ are used to set the electron number in each dot. The device was tuned to the single-electron regime. Gates $P_{1}$, $P_{2}$, $P_{3}$ and $D_{2}$ were connected to homebuilt bias-tees ($RC$=~470 ms), enabling application of d.c. voltage bias as well as high-frequency voltage excitation to these gates. The microwaves were generated using a HP83650A connected to $P_{2}$ via a homemade bias-tee at room temperature. Voltage pulses to the gates were applied using a Tektronix AWG5014. RF reflectometry of the SD was performed using an LC circuit matching a carrier wave of frequency 111.11 MHz. The inductor is formed from a microfabricated NbTiN superconducting spiral inductor with an inductance of 3.0~$\mu$H. The power of the carrier wave arriving at the sample was estimated to be -103 dBm. The carrier signal is only unblanked during read-out. The reflected signal was amplified using a cryogenic Weinreb CITLF2 amplifier and subsequently demodulated using homebuilt electronics. Real time data acquisition was performed using a FPGA (field-programmable gate array DE0-Nano Terasic) programmed to detect tunnel events using a Schmitt trigger. 

\section{Detailed information of the applied pulse sequence}
\begin{figure}[!h]
	\centering
	\begin{tabular}{l l}
		\textbf{a} & \textbf{b} \\
		\includegraphics[width=0.5\textwidth]{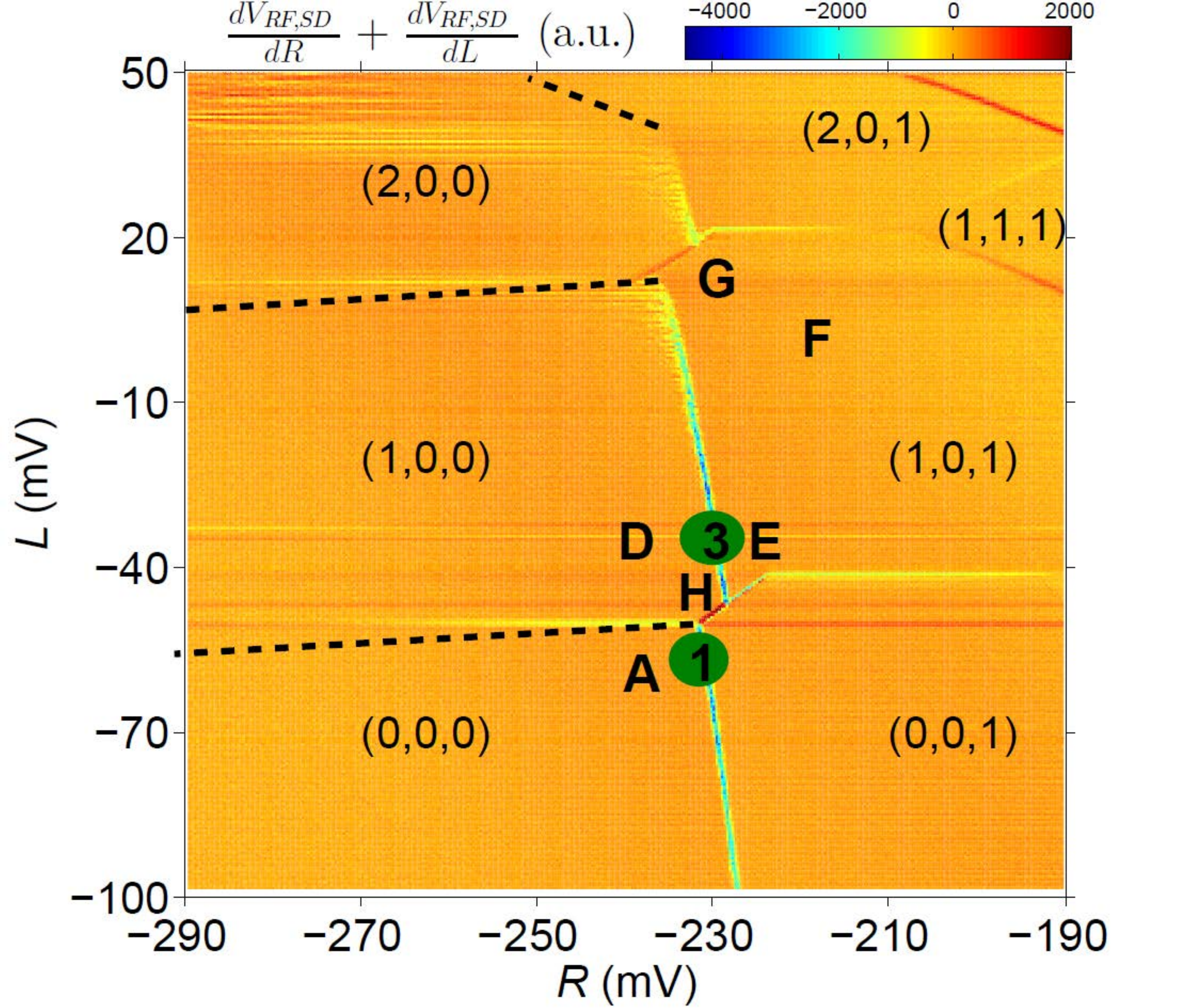} & 		\includegraphics[width=0.5\textwidth]{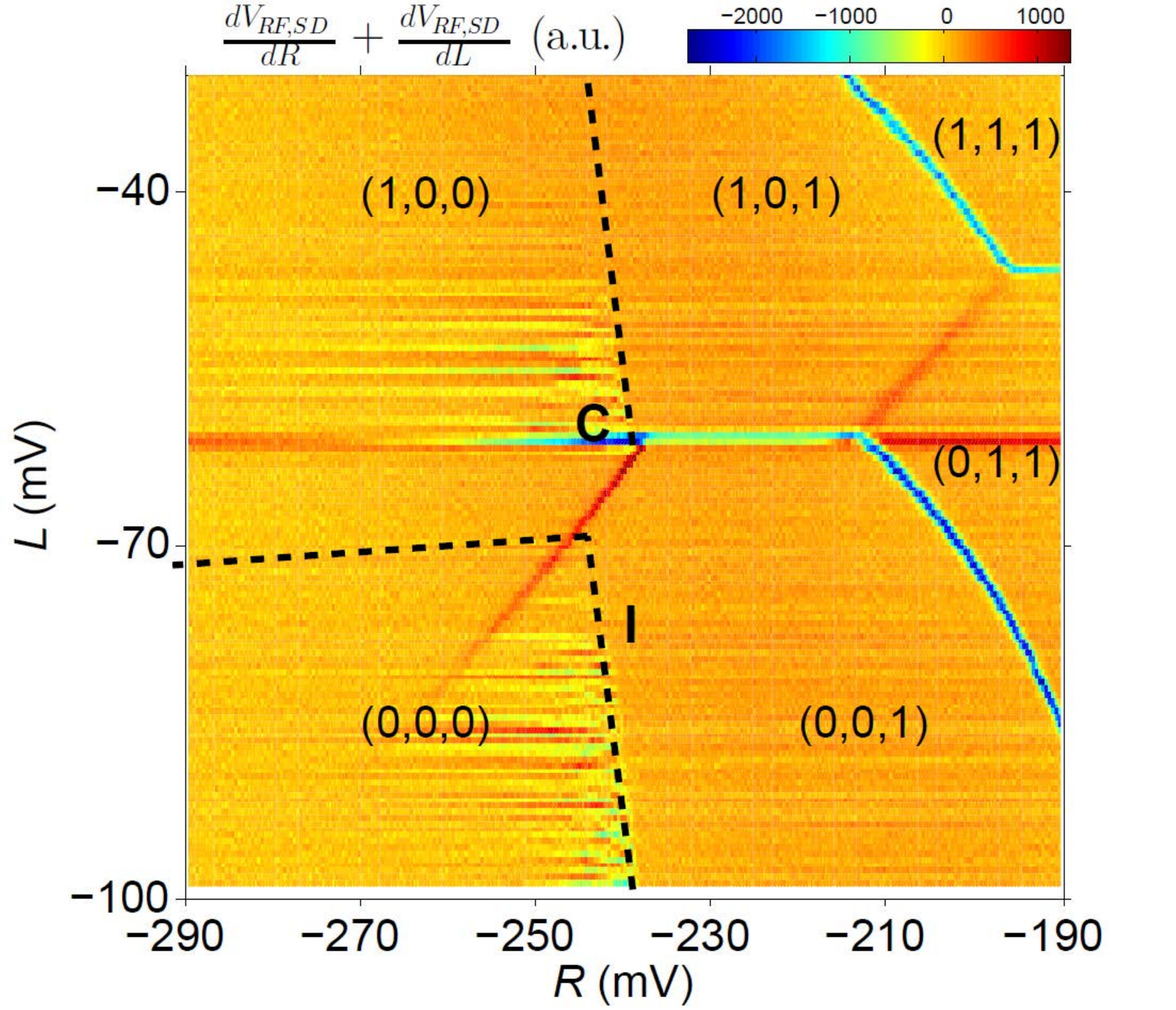} \\
	\end{tabular}
	
	\caption{Charge stability diagram for \textbf{a} $M$=~-412 mV and \textbf{b} $M$=~-382 mV with added details for the pulse sequence as explained in Table S~\ref{tab:details_pulse_sequence}.}
	\label{fig:honeycombs_detailed_pulse_sequence} 
\end{figure}

In this section we give detailed information on the applied pulse sequences. Fig.~S\ref{fig:honeycombs_detailed_pulse_sequence} shows the same charge stability diagram as {\color{black}Fig.~1c} from the main text with additional labeling. Table S~\ref{tab:details_pulse_sequence} gives an explanation including the details of the relevant stages. To correct for slow variations in the dot levels as a function of time (hours timescale), we always calibrate the two read-out stages before each longer measurement such as a complete $T_{1}$ decay ($\sim$ 20,000 datapoints taken after one calibration run, which takes about half an hour). 

\begin{table}[h!]
	\begin{tabular}{ |l | p{13cm} |}
		\hline
		\textbf{Stage(s)} & \textbf{Purpose and details} \\ \hline
		A & Emptying stage (lasting 200~$\mu$s) during which all dots are emptied. \\ \hline
		A $\rightarrow$ Stage 1 & Loading an $\uparrow$-spin from the right reservoir into the right dot. The electrochemical potential of the dot is tuned such that only $\uparrow$-states will enter the dot. We wait 200~$\mu$s in stage 1. \\ \hline 
		Stage 1 $\rightarrow$ C &  Shuttle electron from the right to the left dot via a second-order tunnel process across the middle dot. We temporarily pulse $M$ 30 mV more positive (-382 mV) to increase the long-range shuttling rate. Waiting time at C is 5 $\mu$s. We did not observe any charge exchange with the reservoir at point C. This is also substantiated by the high spin-up fidelity: charge exchange results in the loading of random spins and would therefore increase the offset of the $T_{1}$-curves. \\ \hline
		C $\rightarrow$ D &  We pulse $M$ back to the value $M$=~-412. Waiting time at D is 10~$\mu$s. \\ \hline	
		D $\rightarrow$ E &  Load a random electron from the right reservoir. Waiting time at E is 100~$\mu$s. \\ \hline
		E $\rightarrow$ F &  Prepare exchange gate. Waiting time at F is 1~$\mu$s. \\ \hline
		F $\rightarrow$ G &  Pulse into exchange position G for a varying time. \\ \hline
		G $\rightarrow$ F &  Start read-out sequence. Waiting time at F is 1~$\mu$s. \\ \hline
		F $\rightarrow$ Read-out stage 3 &  Gate voltage pulses larger than $\sim$2 mV produce a spike in the RF-read-out signal. To prevent false events, we therefore first pulse to a position close to the read-out stage (2 mV more positive in $R$) and wait for 2~$\mu$s. Only then, the RF-signal is unblanked and next we pulse into the read-out  configuration for 150~$\mu$s. \\ \hline		
		Read-out stage 3 $\rightarrow$ H & Empty the right dot for 50~$\mu$s. \\ \hline
		H $\rightarrow$ I & Shuttle electron from the left to the right dot. We temporarily pulse $M$ 30 mV more positive to increase the long-range shuttling rate. Waiting time at I is 5~$\mu$s. Point I is positioned relatively far away from the interdot transition (1,0,0)-(0,0,1) to circumvent so-called hot-spots where spin relaxation occurs
		on the sub-$\mu$s timescale~\cite{Srinivasa2013_sup}. We did not observe any charge exchange with the reservoir at point I.\\ \hline
		I $\rightarrow$ Read-out stage 1 & Similar as Read-out stage 3. \\ \hline
		Compensation stage & At the end of the pulse sequence we add a compensation stage of $\sim$0.5 ms to ensure that the total DC-component of the pulse is zero. This prevents unwanted offsets of the dot levels due to the
		bias tees. \\ \hline 				
	\end{tabular}
	\caption{Detailed explanation of the applied pulse sequence as described by Fig.~S\ref{fig:honeycombs_detailed_pulse_sequence}. Unless noted otherwise, we always ramp from one point to the other (ramp time is 1~$\mu$s). The total duration of this sequence is around 900~$\mu$s, not including the compensation stage at the end.}
	\label{tab:details_pulse_sequence}
\end{table}

\clearpage

\section{Virtual gates $L$, $M$ and $R$}
To control the electron number inside each dot we change the voltages on gates $P_{1}$, $P_{2}$ and $P_{3}$. In practice each gate couples capacitively to all three dots. Changing for example $P_{1}$ which couples mostly to dot 1, will therefore also influence dot 2 and 3. To make selective control of each dot easier it is convenient to correct for this capacitive cross-coupling. This can be done by measuring the cross-capacitance matrix for the three gates recording the influence of each gate on every dot. Inverting this matrix allows us to create charge stability diagrams with vertical and horizontal charge transitions in a so-called `virtual gate'-space. In such a `virtual gate'-space, the real gates $P_{1}$, $P_{2}$ and $P_{3}$ have been replaced by linear combinations of the three gates which allow the user to change the chemical potential in one dot, without changing the chemical potential in the neighboring dots.\\

For this experiment the exact relation between  $P_{1}$, $P_{2}$ and $P_{3}$ and the `virtual gates' $L$, $M$ and $R$ used is given by
\[ \left(\begin{array}{c} L \\ M \\R \\ \end{array}  \right) = \left( \begin{array}{ccc}
1.56 & 0.0  & 0.40  \\
0.0  & 1.56 & 0.0 \\
0.20 & 0.51 & 1.05 \end{array} \right) \left(\begin{array}{c} P_{1} - P_{1,\mathrm{offset}} \\ P_{2} - P_{2,\mathrm{offset}} \\ P_{3}- P_{3,\mathrm{offset}}\\ \end{array}  \right)    \] 

with  $ P_{1,\mathrm{offset}} = 23.99$ mV, $ P_{2,\mathrm{offset}} = 61.01$ mV, $ P_{3,\mathrm{offset}} = -111.6 $ mV. 
This set of virtual gates did not perfectly correct for cross-capacitances, as the capacitances themselves turn out to slightly vary with the gate voltages. Working with virtual gates does however give one large freedom to take slices at any angle through the 3D honeycomb diagram that satisfy the requirements for the experiment.\\

\section{Details on calculation of the read-out fidelities}
\begin{figure*}[h!]
	\centering
	\includegraphics[width=0.7\textwidth]{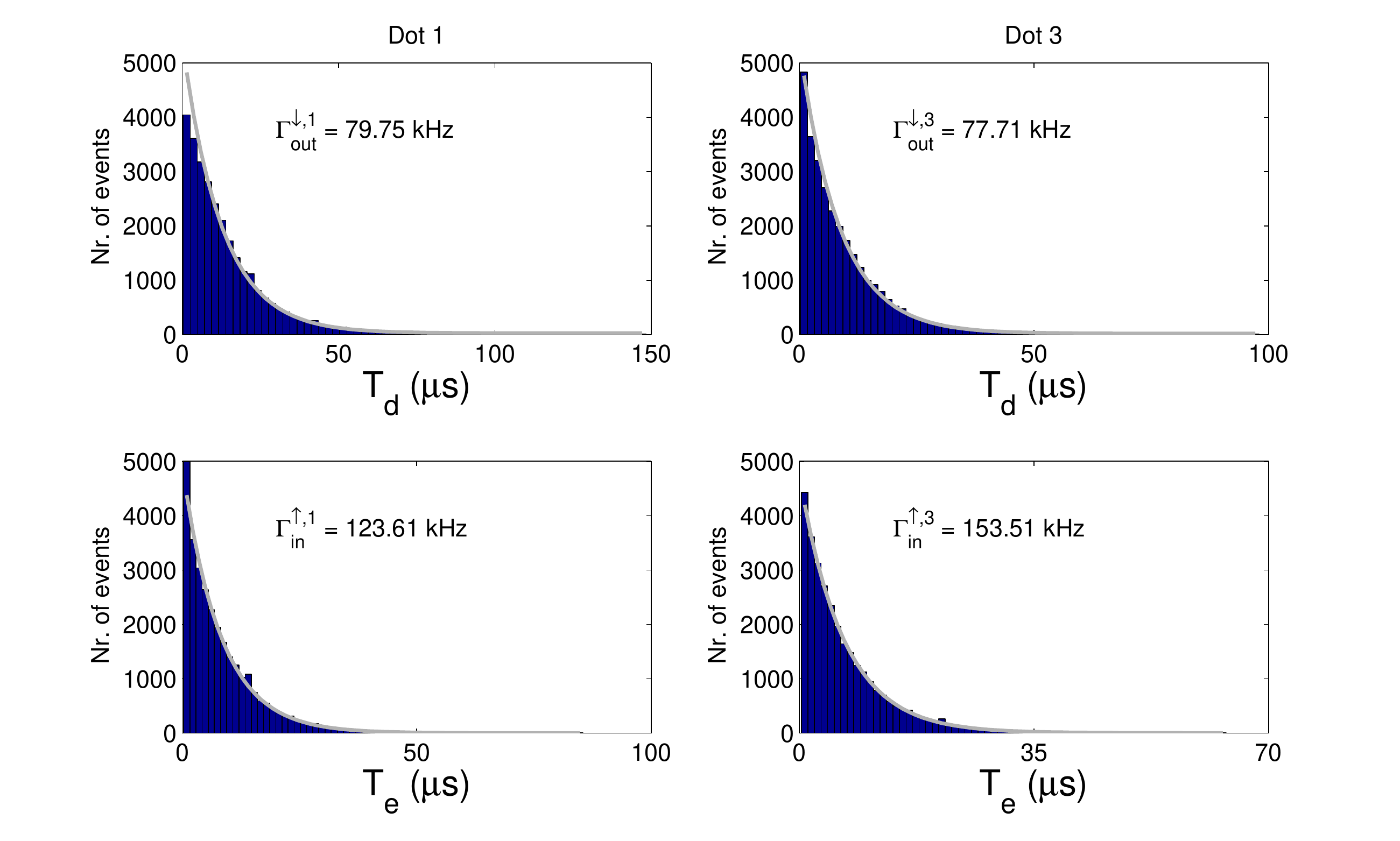}
	\caption{Top row: histograms showing the distribution of the time $T_{d}$ it takes a spin-down electron to tunnel out at the read-out position of each respective dot for a measurement similar to the one shown in {\color{black}Fig.~1d} of the main text. The line is an exponential fit from which we determine the decay rate given by $\Gamma_{out}^{\downarrow,j}+\frac{1}{T_{1}^{j}}$. Bottom row: histograms showing the distribution of the time $T_{e}$ it takes a spin-up electron to tunnel back into the empty dot. The grey line is an exponential fit from which we can extract the tunnel rate given by $\Gamma_{in}^{\uparrow,j}$.}
	\label{fig:tunnel_rates_measurement}
\end{figure*}

\begin{figure*}[h!]
	\centering
	\includegraphics[width=0.5\textwidth]{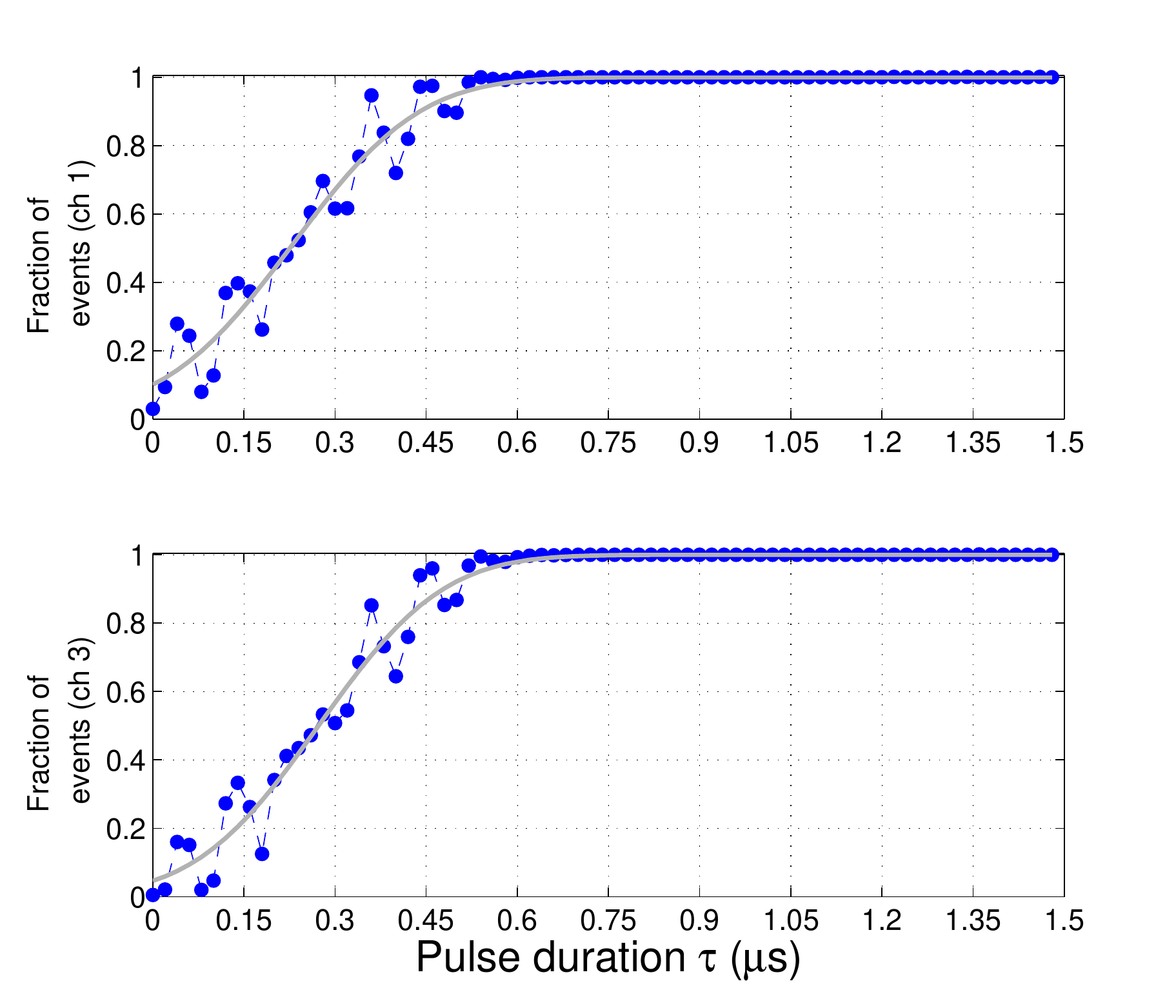}
	\caption{The probability of detecting a pulse with duration $\tau$ for each of the two read-out channels ($B^{j}(\tau)$). Every read-out stage is monitored by a separate input channel of the FPGA. The output of the demodulation box is low-pass filtered at 1 MHz. The sensitivity of the SD is slightly different for the two different read-out stages as they occur at different detunings of the plunger gates. Each datapoint is an average of 999 measurements.}
	\label{fig:bandwidth_measurement}
\end{figure*}

We obtained our read-out fidelities in a similar way as described in~\cite{Baart2015_sup}. The numbers used to obtain these fidelities are summarized in Table S~\ref{tab:overview_parameters_fidelities}. Parameters $T_{1}^{j}$ and $\alpha^{j}$ were extracted from {\color{black}Fig.~1d} of the main text by fitting it to $p^{j} \cdot e^{-t/T_{1}^{j}} + \alpha^{j} $. The tunnel rates with the reservoir, $\Gamma_{out}^{\downarrow,i}$ and $\Gamma_{in}^{\uparrow,i}$, were determined from the histograms depicted in Fig. S~\ref{fig:tunnel_rates_measurement}. The measurement bandwidth, $B^{j}(\tau)$, is extracted from Fig. S~\ref{fig:bandwidth_measurement}. 

\begin{table}[h!]
	\begin{tabular}{ |c |c | c |}
		\hline
		dot nr. & Spin-down fidelity (\%) (worst,best) & Spin-up fidelity (\%) (worst,best) \\ \hline
		1 & 95.8 (95.4, 96.1)  	& 97.7 (97.5, 97.9) \\ \hline
		3 & 96.0 (95.5, 96.3)  	& 98.3 (98.1, 98.5) \\ \hline
	\end{tabular}
	\caption{Read-out fidelities per dot. Values in brackets show the 95\% confidence intervals based on the fits.}
	\label{tab:overview_fidelities}
\end{table}

\begin{table}[h!]
	\begin{tabular}{ |c | c |c | c | c |c | c|}
		\hline
		dot nr. & $T_{1}$ (ms) (worst,best) & $\Gamma_{out}^{\downarrow,i}$ (kHz) (worst,best) & $\Gamma_{in}^{\uparrow,i}$  (kHz) (worst,best) & Read-out time ($\mu$s) & $T_{wait}^{3}$ ($\mu$s) \\ \hline
		1 & 17.8 (17.2, 18.4) & 79.75 (78.89, 80.61) & 123.61 (125.29, 121.92) & 150 & 60.25 \\ \hline
		3 & 17.7 (17.2, 18.3) & 77.71 (76.93, 78.48) & 153.51 (154.51, 152.51)& 150 & - \\ \hline
	\end{tabular}
	\caption{Overview of the required parameters to calculate the read-out fidelities as described in~\cite{Baart2015_sup}. Values in brackets show the 95\% confidence intervals based on the fits.}
	\label{tab:overview_parameters_fidelities}
\end{table}

\clearpage
\section{Simulation of the experiment}
\label{sec:Simulation}
In this section we describe how we calculated the simulated data as shown in {\color{black} Fig.~3b} of the main text. 

\subsection{Theoretical model}
We model our system using the following hamiltonian, defined in the $S(200)$, $S(110)$, $T_{0}(110)$, $S(101)$, $T_{0}(101)$ basis. 
\[ H = \left( \begin{array}{ccccc}
- \epsilon /2 	& 	t_{l}  				& 0 				& 	0 				& 0  \\
t_{l}  			& -\delta 				& \Delta E_{z,21}/2   &   t_{r}			& 0 \\
0 				&  \Delta E_{z,21}/2		& -\delta			& 	0				& t_{r} \\
0 				&  t_{r}				& 0					& \epsilon /2		& \Delta E_{z,31}/2 \\
0 				&  0					& t_{r}				& \Delta E_{z,31}/2	& \epsilon /2 \end{array} \right)  \] 

The matrix elements are defined as follows: 
\begin{itemize}
	\setlength\itemsep{0.0em}
	\item $\epsilon$ is the detuning between the outer dots
	\item $\delta$ is the detuning of the middle dot with respect to the average of the outer dots
	\item $t_{l}$ is the tunnel coupling between the $S(200)$ and $S(110)$ state 
	\item $t_{r}$ is the tunnel coupling between the (110) and (101) state
	\item $\Delta E_{z,21}$ is the difference in Zeeman energy between dot 1 and 2, defined as $E_{z,2}-E_{z,1}$.
	\item $\Delta E_{z,31}$ is the difference in Zeeman energy between dot 1 and 3, defined as $E_{z,3}-E_{z,1}$.
\end{itemize} 
We have to solve the time-dependent Schr\"odinger equation to incorporate the finite rise- and falltime of the AWG:
\begin{equation}
	i \hbar \frac{\partial}{\partial t}  \Psi(t) = H\left(\epsilon(t),\delta(t)\right) \Psi(t)
\end{equation}
We numerically approximate the time evolution of $\Psi(t)$ by discretizing time in steps $\Delta t$ and calculating\\ ${\Psi(t+\Delta t) = e^{-i H(t) \Delta t / \hbar} \Psi(t)}$. We model the time dependence of $\epsilon(t)$ and $\delta(t)$ as follows:\\

If $t < t_{f}$:
\begin{equation}
	\epsilon(t)=\left[\epsilon(t_{f})-\epsilon(0)\right]\cdot (1-e^{-t/t_{rise}}) + \epsilon(0)
	\label{eq:risetime1}
\end{equation}
\begin{equation}
	\delta(t)=\left[ \delta(t_{f})-\delta(0)\right]\cdot (1-e^{-t/t_{rise}}) + \delta(0)
\end{equation}
Else: 
\begin{equation}
	\epsilon(t)=\left[ \epsilon(t_{f})-\epsilon(0)\right]\cdot (1-e^{-t_{f}/t_{rise}})\cdot e^{-(t-t_{f})/t_{rise}}  + \epsilon(0)
\end{equation}
\begin{equation}
	\delta(t)=\left[\delta(t_{f})-\delta(0)\right]\cdot (1-e^{-t_{f}/t_{rise}}) \cdot e^{-(t-t_{f})/t_{rise}} + \delta(0)
	\label{eq:risetime4}
\end{equation}

Where $(\epsilon(0)$, $\delta(0))$ denotes the starting point (red star symbol in {\color{black}Fig.~2b}), $(\epsilon(t_{f})$, $\delta(t_{f}))$ the detuning point programmed in the AWG, $t_{f}$ the programmed time to spend in $(\epsilon(t_{f})$, $\delta(t_{f}))$, and $t_{rise}$ the risetime of the pulse arriving at the sample.

\begin{figure*}[h!]
	\centering
	\includegraphics[width=0.5\textwidth]{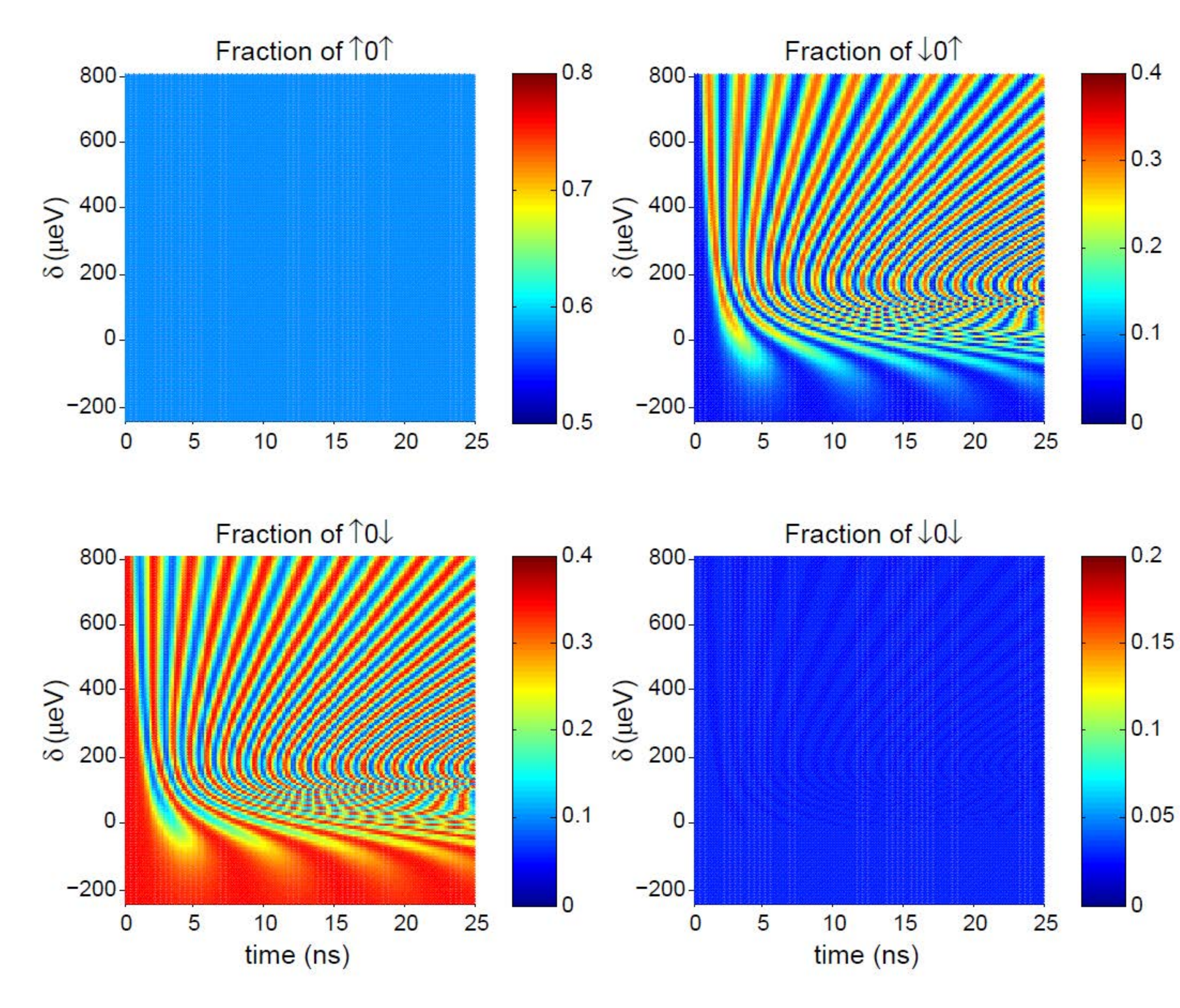}
	\caption{Same simulation as {\color{black}Fig.~3a} of the main text without the dephasing term.}
	\label{fig:simulation_without_dephasing}
\end{figure*}

We add pure electrical dephasing by multiplying with an amplitude decay of the form $e^{-\left(t_{\mathrm{SWAP}}/T_{2}^{*} \right)^{2}}$~\cite{Dial2013_sup}. The sensitivity of $T_{2}^{*}$ to charge noise is approximated by $T_{2}^{*} = \frac{T_{2,\mathrm{factor}}}{\sqrt{\left( \left(\frac{\partial}{\partial \epsilon} J(\epsilon(t_{f}),\delta(t_{f})) \right)^{2} + \left(\frac{\partial}{\partial \delta} J(\epsilon(t_{f}),\delta(t_{f})) \right)^{2}  \right)} }$. The simulations shown use $T_{2,\mathrm{factor}}=$ 0.25~ns. This model does not incorporate the change in $T_{2}^{*}$ during the risetime of the pulse, which could be incorporated in future work. Especially pulses traversing the transition from superexchange (1,0,1) to nearest neighbour exchange (1,1,0) suffer from a temporarily shorter $T_{2}^{*}$ than the value calculated at $(\epsilon(t_{f}),\delta(t_{f}))$. For completeness Fig. S~\ref{fig:simulation_without_dephasing} shows the simulation with no dephasing.

We correct for the read-out fidelities by introducing the fidelity matrix $\textbf{F}$:
\[ \mathrm{\textbf{F}} = \left( \begin{array}{cccc}
F_{\uparrow}^{1}F_{\uparrow}^{3}  	& 	F_{\uparrow}^{1}(1-F_{\downarrow}^{3}) & (1-F_{\downarrow}^{1})F_{\uparrow}^{3} & 	(1-F_{\downarrow}^{1})(1-F_{\downarrow}^{3}) 			 \\
F_{\uparrow}^{1}(1-F_{\uparrow}^{3}) &  F_{\uparrow}^{1}F_{\downarrow}^{3}	& (1-F_{\downarrow}^{1})F_{\uparrow}^{3}   &   (1-F_{\downarrow}^{1})F_{\downarrow}^{3}		 \\
(1-F_{\uparrow}^{1})F_{\uparrow}^{3} &  (1-F_{\uparrow}^{1})(1-F_{\downarrow}^{3}) &  F_{\downarrow}^{1} F_{\uparrow}^{3}	& 	F_{\downarrow}^{1}(1-F_{\downarrow}^{3}) \\
(1-F_{\uparrow}^{1})(1-F_{\uparrow}^{3}) & (1-F_{\uparrow}^{1})F_{\downarrow}^{3}	& F_{\downarrow}^{1} (1-F_{\uparrow}^{3}) & F_{\downarrow}^{1}F_{\downarrow}^{3}	 \end{array} \right)  \] 
Where $F_{\uparrow}^{i}$ and $F_{\downarrow}^{i}$ are the read-out fidelities for dot $i$ (see Table S~\ref{tab:overview_fidelities}). We denote the measured four two-spin probabilities at time $t$ by $\Pi_{\mathrm{measured}}(t) = (P_{\uparrow\uparrow},P_{\uparrow\downarrow},P_{\downarrow\uparrow},P_{\downarrow\downarrow})$. The actual two-spin probabilities $\Pi(t)$, i.e. corrected for the read-out fidelities, can be found by solving the linear equations $\Pi_{\mathrm{measured}}(t) = \mathrm{\textbf{F}} \Pi(t)$. The linear equations are solved by numerically minimizing the squared length of the vector $\Pi_{\mathrm{measured}}(t) - \mathrm{\textbf{F}} \Pi(t)$ and restricting the solutions $\Pi(t)$ to physical solutions (all elements have to be positive and the sum of the elements has to be one, since the vector consists of four two-spin probabilities). Using this method we determine $\Pi(t=0)$, which is used as initial condition for the simulation. The outcome of the simulation, $\Pi(t)$, is multiplied by \textbf{F} to calculate the theoretically expected $\Pi_{\mathrm{measured}}(t)$ as shown in {\color{black}Fig.~3b} of the main text.

\subsection{Input parameters}

\textbf{Zeeman energy $E_{z}$:} The Zeeman energy in each dot has been measured using electric-dipole spin resonance (EDSR) in a similar way as in Ref.~\cite{Shafiei2013_sup}. At 3.2~T the Zeeman energy is 19.380 GHz, 19.528 GHz and 19.510 GHz for dot 1, 2 and 3 respectively. 

\begin{figure}[!h]
	\centering
	\begin{tabular}{l l}
		\textbf{a} & \textbf{b} \\
		\includegraphics[width=0.5\textwidth]{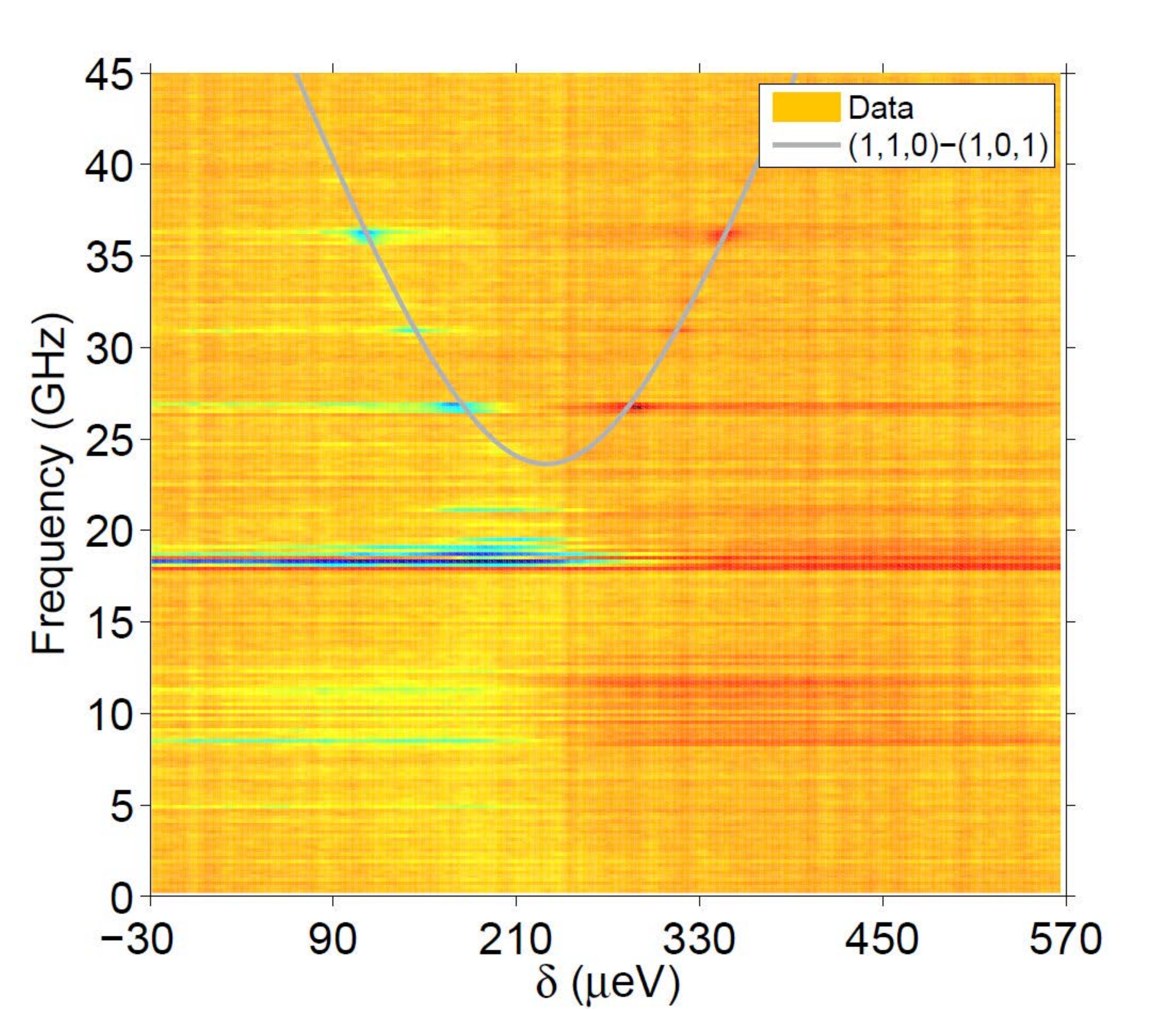} & 		\includegraphics[width=0.5\textwidth]{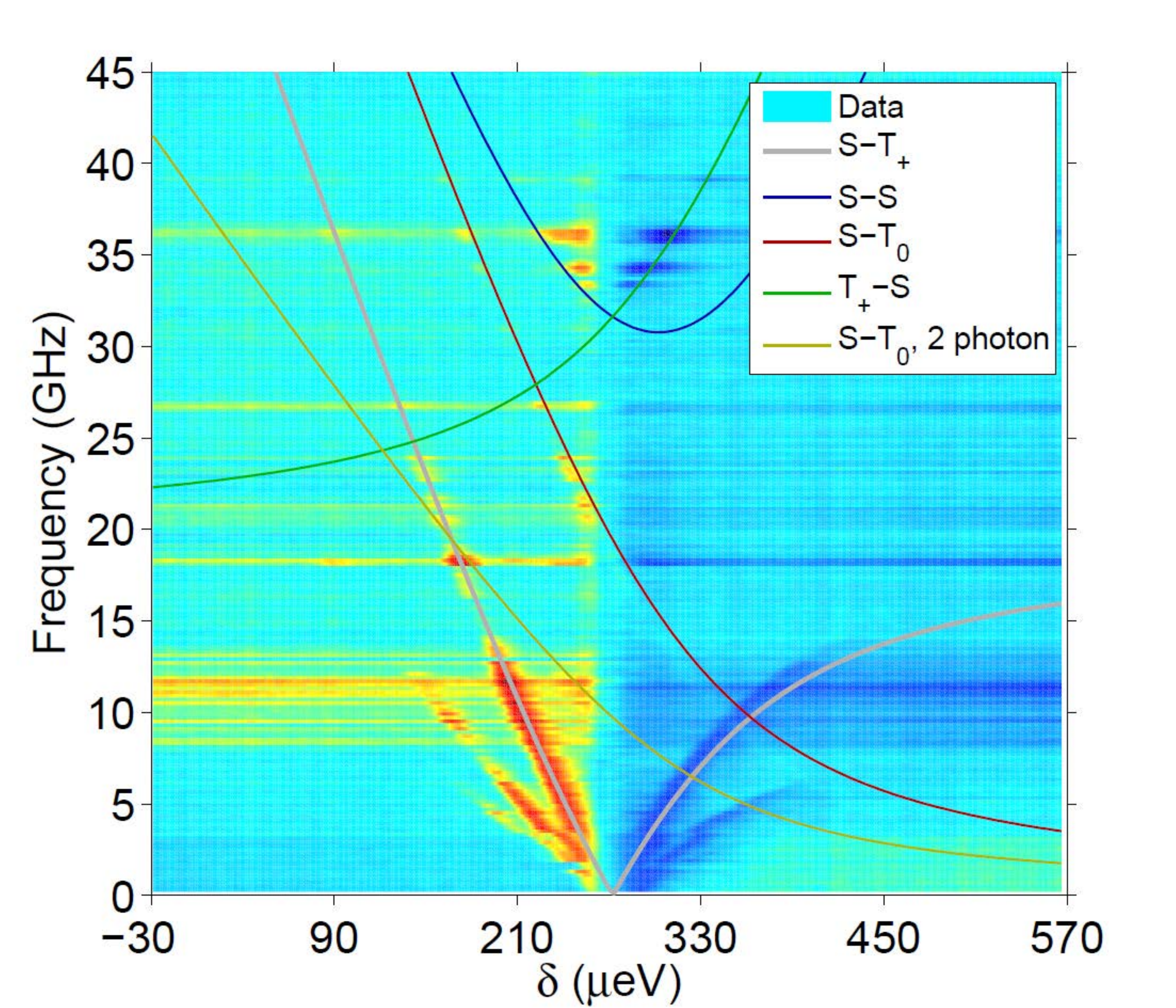} \\
	\end{tabular}
	
	\caption{Charge sensor signal as a function of frequency versus $\delta$ between: \textbf{a} (1,1,0) and (1,0,1) at $\epsilon=-230~\mu$eV, \textbf{b} (2,0,0) and (1,1,0) at $\epsilon=230~\mu$eV. The data in \textbf{a} is fitted to $\sqrt{((M-M_{\mathrm{offset}}) \cdot \alpha_{M})^2+4t_{r}^{2}}$ giving $t_{r}$ = 11.8 $\pm$ 0.6 GHz. In \textbf{b} we fit the $S-T_{+}$-transition that is clearly visible to $\left| \left(\sqrt{((M-M_{\mathrm{offset}}) \cdot \alpha_{M})^{2}+4t_{l}^{2}}-((M-M_{\mathrm{offset}}) \cdot \alpha_{M})  \right)/2 - \left(E_{z,1}+E_{z,2}\right)/2 \right|$  giving $t_{l}$ = 15.4 $\pm$ 1.3 GHz. Based on this fit we also plot the expected position of multiple other transitions.}
	\label{fig:PAT} 
\end{figure}

\textbf{Tunnel couplings $t_{l}$ and $t_{r}$:} The tunnel coupling at zero detuning between neighbouring dots was measured using photon-assisted tunneling~\cite{Oosterkamp1998a_sup}, see Fig.~S~\ref{fig:PAT}. These measurements were also used to determine the lever arm $\alpha_{i}$ of each gate for $\epsilon$ and $\delta$. $t_{r}$ was measured at: $(\epsilon,\delta)= (-230,180)$~$\mu$eV. This is close to the detuning where the data of {\color{black}Fig.~3a} was acquired. The value of $t_{l}$ had to be measured however for a different tuning: $(\epsilon,\delta)= (230,180)$~$\mu$eV. Tunnel couplings depend strongly on the exact gate voltages. We measured for example that decreasing $\delta$ by 300~$\mu$eV whilst staying on the (2,0,0)-(1,1,0) transition decreases $t_{l}$ by $\sim$30\%. At the detuning of the experiment most datapoints are taken away from the (2,0,0)-(1,1,0) transition in the direction of the (1,1,0)-regime where $t_{l}$ will be even further reduced~\cite{Medford2013_sup}. We find a good match between experiment and simulation by keeping $t_{r}$ fixed at the measured 11.8~GHz and lowering the measured $t_{l}$ by 45\% to 8.5~GHz. 
We have verified that changing $t_{l}$ is the most efficient way to get good agreement with experiment. $t_{l}$ strongly influences the oscillations for positive $\delta$, whilst the product of $t_{l}$ and $t_{r}$ dominates for negative $\delta$. This allowed us to determine that $t_{l}$ had to be changed and not $t_{r}$. Future work could incorporate the precise dependence of tunnel coupling on detuning.

\begin{figure*}[h!]
	\centering
	\includegraphics[width=0.9\textwidth]{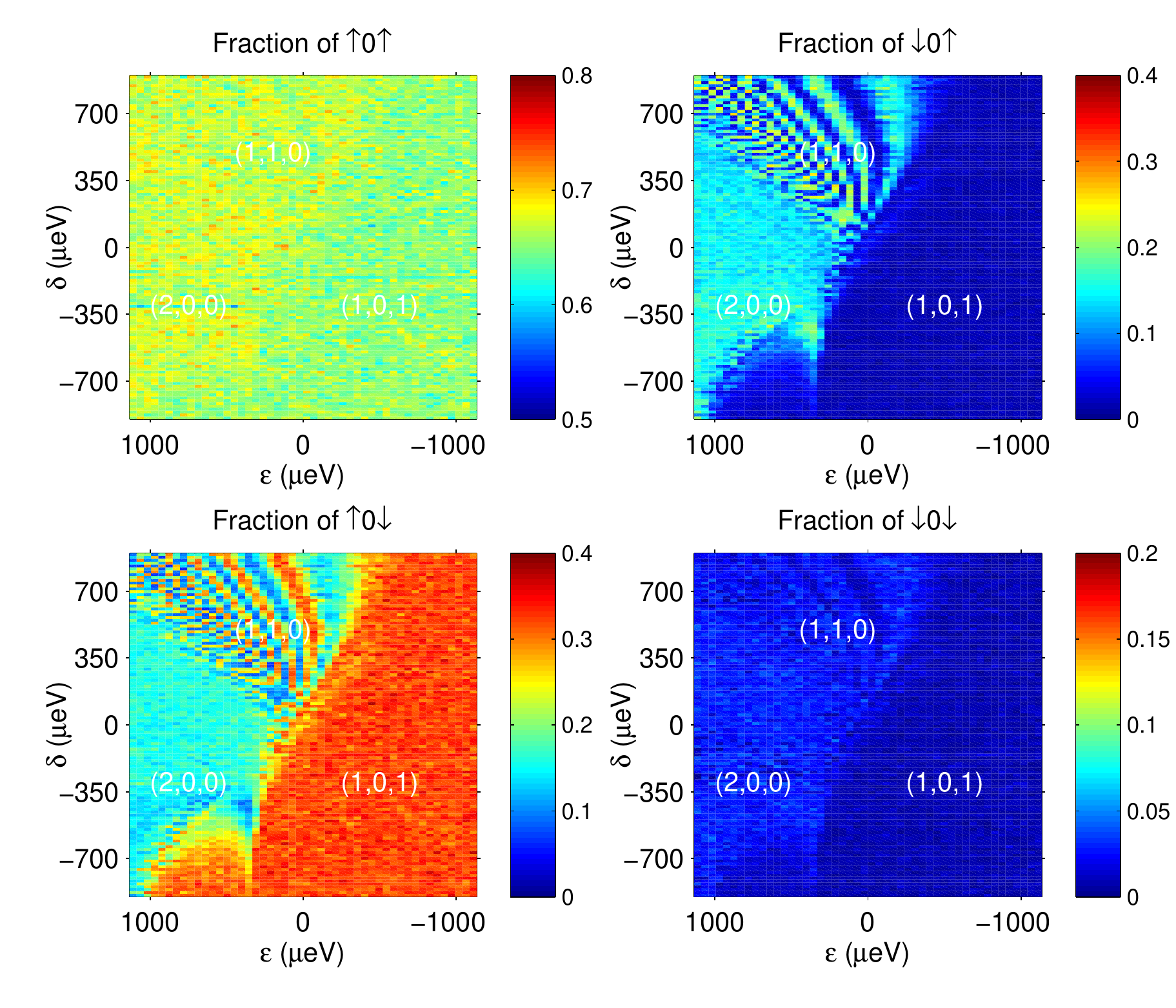}
	\caption{Pulsed version of charge stability diagram {\color{black}Fig.~2b} of the main text. We create a mixture of $\uparrow0\downarrow$ and $\uparrow0\uparrow$ in (1,0,1) and pulse for 5~ns to the same detunings as {\color{black}Fig~2b} and detect the spin-state.}
	\label{fig:pulsed_Mercedes_plot}
\end{figure*}

\textbf{$\epsilon(t_{f})$, $\delta(t_{f})$ and $t_{rise}$ :} The model described in Eq.~(\ref{eq:risetime1})-(\ref{eq:risetime4}) is an approximation to the true shape of the pulse. The measured output of the AWG including the low pass filter (MiniCircuits SBLP-300+) can be modeled well with $t_{rise} \approx 0.8$~ns. The coax cables inside the dilution refrigerator show a shorter $t_{rise}$ ($<$600 ps), but in addition suffer from the skin effect. Due to this effect it takes several ns to reach the final voltage value on the sample. In practice this means that for short pulses ($<$ few ns) Eq.~(\ref{eq:risetime1})-(\ref{eq:risetime4}) work fine if $\epsilon(t_{f})$ and $\delta(t_{f})$ are reduced in magnitude to incorporate the skin effect. This is also seen in practice by measuring a pulsed version of charge stability diagram {\color{black}Fig.~2b} of the main text. {\color{black}Fig.~2b} depicts the charge stability diagram when changing the detunings relatively slowly ($\sim$ms timescale per horizontal trace). To verify that the fast ($\sim$ns) pulses arrive at slightly different points in detuning-space, we create a mixture of $\uparrow0\downarrow$ and $\uparrow0\uparrow$ in (1,0,1) and pulse for 5~ns to the same detunings as {\color{black}Fig~2b} and detect the spin-state. This pulse starts at $(\epsilon,\delta)\approx (-950,0)$. Fig.~S\ref{fig:pulsed_Mercedes_plot} shows the expected three  regimes: (1) as long as the pulse stays within the (1,0,1)-region no spin evolution occurs, (2) pulsing into the (2,0,0)-region leads to a mixture of the $\uparrow0\downarrow$, $\downarrow0\uparrow$ and $\uparrow0\uparrow$ states as a consequence of full dephasing during the superexchange and (3) pulsing into (1,1,0) induces coherent nearest-neighbour exchange. 
Comparing Fig.~S\ref{fig:pulsed_Mercedes_plot} and {\color{black}Fig~2b} shows that the transition from (1,0,1) to (2,0,0) takes place at a more positive $\epsilon$ as expected: $\sim$300~$\mu$eV. This is attributed to the skin effect inside the coax cables. A similar measurement for a pulse time of 1~$\mu$s does not show this shift. The data in {\color{black}Fig.~3a} is taken for $\epsilon(t_{f})=0$. To compensate for the skin effect we need to input a more negative value for $\epsilon$ in the simulation to get good agreement with the data: -170 $\mu$eV. To make a perfect model one could first measure the precise pulse shape arriving at the sample more carefully.\\
Despite some uncertainty in the exact value of some of the input parameters, within a reasonable range of inputs all give a similar qualitative picture. They all show the change from a regime driven by $\Delta E_{z,31}$ for negative $\delta$, to superexchange oscillations that become faster as $\delta$ is increased to a regime of nearest-neighbour oscillations that slow down as $\delta$ becomes more positive. This underlines that our experimental data indeed shows the transition from superexchange to nearest-neighbour exchange. 

\clearpage

\section{Comparison of the quality factor of superexchange versus nearest-neighbour exchange}
In this section we make a comparison of the charge-noise sensitivity of superexchange versus nearest-neighbour exchange. This allows us to compare the quality factor, $Q$, in the regime where charge noise is the dominant source of dephasing (i.e. $J> \Delta E_{Z}$). $Q$ is proportional to the number of visible exchange oscillations within the dephasing time: ${Q \propto J \cdot T^{*}_{2}}$.\\

We start with an approximation of the nearest-neighbour exchange interaction $J_{N}$ by~\cite{Loss1998a_sup}
\begin{equation}
	J_{N}\approx - \frac{t^{2}_{N}}{\epsilon},
\end{equation}
with $t_{N}$ the neareast-neighbour tunnel coupling and $\epsilon$ the detuning between the dots. This approximation is valid in the regime where $t_{N} \ll |\epsilon|$ and $\epsilon <  0$. We assume that fluctuations in $\epsilon$ are the dominant contribution to dephasing and neglect dephasing from other fluctuating parameters~\cite{Dial2013_sup}. The sensitivity to charge noise is then given by
\begin{equation}
	\frac{\partial}{\partial \epsilon} J_{N} = \frac{t_{N}^{2}}{\epsilon^{2}}
\end{equation}
For a fixed $J_{N}$ and $t_{N}$ the sensitivity to charge noise is
\begin{equation}
	\frac{\partial}{\partial \epsilon} J_{N} = \left(\frac{J_{N}}{t_{N}}\right)^{2}
	\label{eq:chargenoiseshortrange}
\end{equation}
This results in a $Q$ of 
\begin{equation}
	Q \propto \frac{J}{\frac{\partial}{\partial \epsilon} J} = \frac{t_{N}^2}{J_{N}}
	\label{eq:Q_shortrange}
\end{equation}
Note that in this approximation, a larger $Q$ for a fixed $J$ is obtained by increasing the tunnel coupling.\\

\begin{figure*}[h!]
	\centering
	\includegraphics[width=1.0\textwidth]{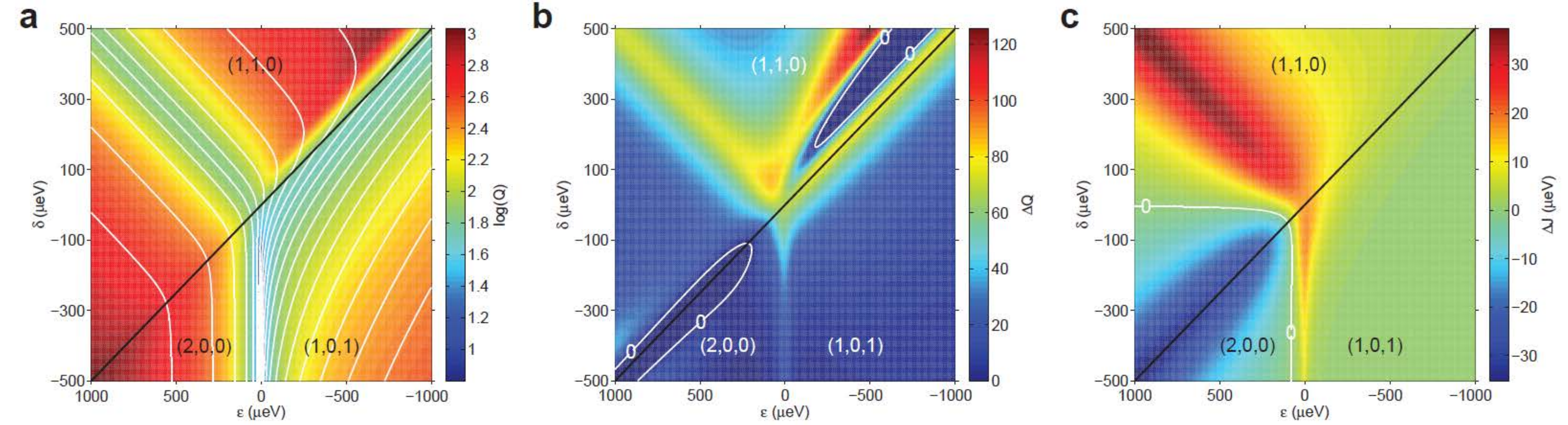}
	\caption{\textbf{a} Numerically approximated $\log(Q)$ of the exchange oscillations as a function of $\delta$ and $\epsilon$ for $t=40~\mu$eV. White lines represent contour plots of fixed $J$ for several different values of $J$. Black line denotes the transition between nearest-neighbour and superexchange. \textbf{b} Change in $Q$ when $t$ is doubled to $t=80~\mu$eV: $\Delta Q = Q(t=80)-Q(t=40)$. White line represents contour plot of $\Delta Q = 0$. \textbf{c} Change in $J$ when $t$ is doubled to $t=80~\mu$eV: $\Delta J = J(t=80)-J(t=40)$. White line represents contour plot of $\Delta J = 0$. The decrease in $J$ in the (2,0,0)-region is caused by the increased anti-crossing between the $T_{0}(1,1,0)$- and $T_{0}(1,0,1)$-state.}
	\label{fig:charge_noise}
\end{figure*}

A similar first order approximation can be made for the superexchange regime. It however only gives reliable results for a small portion of the detuning space. To give a fair comparison between nearest-neighbour and superexchange we therefore used the same model as in section~\ref{sec:Simulation} to numerically calculate $J$ and the effect of charge noise for each value of $\delta$ and $\epsilon$. Assuming uncorrelated, equal magnitude noise in $\delta$ and $\epsilon$ results in the following expression for the charge-noise sensitivity
\begin{equation}
	\sqrt{\left( \frac{\partial}{\partial \epsilon} J \right)^{2} + \left(\frac{\partial}{\partial \delta} J \right)^{2}}
	\label{eq:chargenoiselongrange}
\end{equation}
and we approximate $Q$ by
\begin{equation}
	Q \propto \frac{J}{\sqrt{\left( \frac{\partial}{\partial \epsilon} J \right)^{2} + \left(\frac{\partial}{\partial \delta} J \right)^{2}}} 
	\label{eq:Q_longrange}
\end{equation}
For simplicity we set the $g$-factor differences to 0 and assume equal nearest-neighbour tunnel couplings $t$. The calculated $\log(Q)$ for $t=40~\mu$eV is shown in Fig.~S\ref{fig:charge_noise}a. White contour lines indicate constant $J$ for several different values of $J$. From Fig.~S\ref{fig:charge_noise}a we can conclude the following if we limit ourselves to the (1,1,0)- and (1,0,1)-regime: for a fixed $J$, the best $Q$ is always obtained by operating in the short-range regime (1,1,0). (Please note that all white lines in (1,0,1) eventually enter the (1,1,0)-regime; for several contour lines this happens outside the plotted detuning range of the figure.) 

It is possible to boost the performance of both superexchange and nearest-neighbour exchange by increasing the interdot tunnel coupling. In Fig.~S\ref{fig:charge_noise}b and c we plot how much $Q$ and $J$ change respectively when $t$ is doubled to $t=80~\mu$eV. This shows that the $Q$ and $J$ of superexchange both improve inside the (1,0,1)-region by increasing $t$, allowing one to induce more oscillations at higher speed. Inside the nearest-neighbour region (1,1,0) there is also a large region where both $Q$ and $J$ increase. 

We can also include the (2,0,0)-region in the discussion which can be reached using long-range tunneling. This requires the timing jitter of the tunneling from the (1,0,1)- to the (2,0,0)-region to be small compared to $J$. This condition can be satisfied by making the tunnel couplings sufficiently large. In that case, the contour lines in Fig.~S\ref{fig:charge_noise}a indicate that for a fixed $J$, the $Q$'s are similar for both the nearest-neighbour and superexchange region. This can be understood from the fact that in the far detuned regime, the energy splitting between the $T_{0}$ and $S$ state ($J$) is linearly proportional to the detuning and the tunnel coupling has a negligible influence. We expect the nearest-neighbour and superexchange to show even better $Q$'s when operating in the very far detuned regime inside (2,0,0), where the exchange is set by the orbital splitting and the sensitivity to charge noise is low~\cite{Dial2013_sup}. In that regime we expect the nearest-neighbour and superexchange to have the same quality factor.

As a final observation we note that nearest-neighbour exchange can never fully be turned off and can at most be reduced to $t^{2}/U$ with $U$ the charging energy of the dot~\cite{Loss1998a_sup}. Superexchange can be set closer to zero however as $\delta$ can easily be made larger than $U$: this is beneficial to fully decouple qubits.

\end{document}